# DISCOVERY AND STUDY OF THE ECLIPSING VARIABLE STAR GRIGORIEV 1

*V. S. Grigoriev, D. V. Denisenko\**

State education center "Vorobyovy gory" (Moscow Palace of Pioneers), Moscow, Russia

This work is dedicated to the discovery of the eclipsing variable star Grigoriev 1 and determining the parameters of its binary system. The star was found through the systematic checking of ultraviolet sources of GALEX space observatory in Pegasus constellation. Analysis of the ZTF project data has shown the presence of eclipses with two magnitudes depth and a period of 6.5997 days. The duration of eclipse covers only 1% of the orbital period, and the partial phases are at least 30 times shorter. Those parameters imply that the star is a detached binary system with a white dwarf seen edge-on. The new object was added to the International Variable Star Index AAVSO VSX as Grigoriev 1. Over the 10 million objects in VSX database there are only 188 variable stars of EA/WD type. Grigoriev 1 has the longest orbital period out of those. Moreover, its absolute magnitude M at maximum light from Gaia space observatory data is about +6.8. On the "color-luminosity" diagram it occupies the intermediate position between hot subdwarfs and white dwarfs which makes it even more interesting and worth studying at professional telescopes.

The project "Center of astronomical object discoveries" is being realized at the Center of Astronomical and Space Education of Moscow Palace of Pioneers at Vorobyovy gory since 2021. Over thirty school students have become discoverers of more than 60 supernovae and 120 variable stars during those four years. Most of them are typical representatives of the main variability classes (pulsating, eclipsing, erupting, cataclysmic ones). However, from time to time schoolchildren manage to find the stars with no analogs among the known objects. Yet the number of variable stars discovered by the professional surveys and observatories is growing, which makes the individual search harder and harder each year. The International Variable Star index (AAVSO VSX) database has overtaken 10 million records. This is why one must invent new search methods, including those using the publicly available astronomical resources of the World Wide Web.

The new idea for searching was inspired by the variable star Minkovskiy 6 discovered earlier by the high-school student Artur Minkovskiy on the digitized Palomar plates (Nasonov et al., 2024). This star appeared much brighter in the UV-band than in visible light. We have concluded that interesting objects should be searched for among the ultraviolet sources. The work was started during the astronomical project camp at "Komanda" Education center of Moscow Palace of Pioneers.

The search was performed as follows. Objects which had blue color on GALEX images, but appeared as usual stars on DSS colored images, were visually selected. The light curves of those objects were checked at the Zwicky Transient Facility (ZTF) website (Masci et al., 2019). Some of those objects turned out to be quasars, others – already discovered variable stars. At last, on the 24$^{th}$ of October 2024 the author of this work managed to find the star in Pegasus constellation changing its brightness by 2 magnitudes (Fig. 1) but not listed in any catalogs of variable stars. It was not present either in Simbad astronomical database, nor in the International Variable Star Index AAVSO VSX (Watson et al., 2006). Checking Vizier website has shown that there is no mention of variability in the data of Gaia space observatory, nor in ATLAS, ZTF, CSS, ASAS-SN or any other survey projects. That meant the new variable star was discovered whose brightness drops from 18$^{th}$ to 20$^{th}$ magnitude from time to time.

WinEffect program by V. P. Goranskij was used to determine the period of eclipses. It was found to be 6.5997 days (Fig. 2). The star was submitted to the International Variable Star Index with the name Grigoriev 1, and it was included in catalog as EA/WD (eclipsing binary of Algol type with the white dwarf) less than 12 hours since the discovery. The parameters of the new star are given in Table 1. It turned out to be extremely rare. Out of 10 million objects in VSX database there are only 188 variable stars of EA/WD type. Among them, Grigoriev 1 has the second longest orbital period. Only GSC 04047–00113 in Cassiopeia has the longer period, but the eclipses of white dwarf in this system are caused by the accretion disk around the secondary component and have much smaller depth, merely 0.06 magnitude. Among the detached white dwarf + red dwarf binary systems the variable star Grigoriev 1 is the leader by a margin of 2.5 times from the closest analogue with the period of 2.308 days.

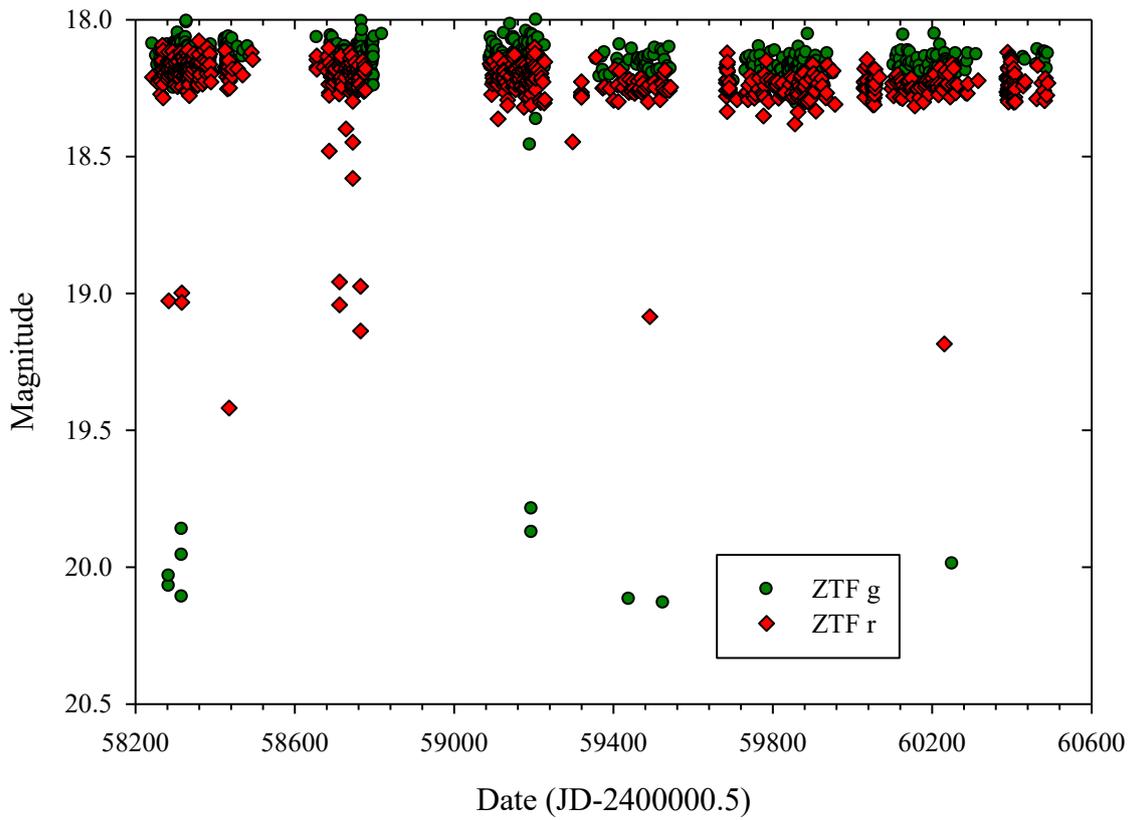

**Fig. 1.** Light curve of the variable star Grigoriev 1 from ZTF project. Green circles – observations in *g* filter, red diamonds – in *r* filter.

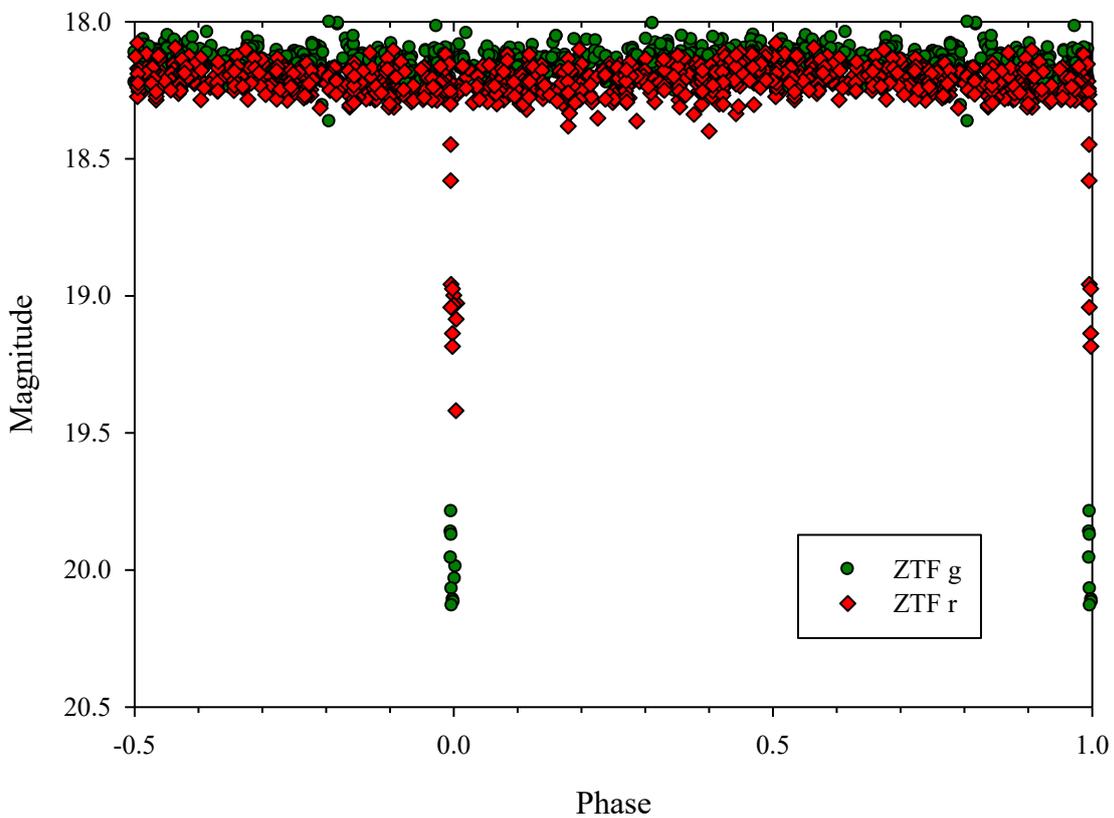

**Fig. 2.** Phase plot of Grigoriev 1 with the 6.5997 d period

The eclipses in Grigoriev 1 binary system last only 1% of the orbital period, that is about 1.5 hours, with the partial phases at least 30 times shorter. That means that the size of the eclipsing component is also about 30 times larger than the size of the star being eclipsed. Since the brightness of star during the eclipse is reduced by about 6 times (2 magnitudes), the eclipsed component is about 5 times brighter than the eclipsing one. This could only be possible if the eclipsing object is cooler (for instance, red dwarf), and the eclipsed component is hotter (white dwarf or subdwarf). That is confirmed by the depth of eclipse in different bands of visible light. In *g* filter the brightness falls by 2 magnitudes (6.3 times), in *r* filter – by 1.4 mag (3.6 times).

Table 1. Observational parameters of the variable star Grigoriev 1

| Position (Epoch 2000.0) | 22 20 52.12 +33 14 51.1 |
|---|---|
| Constellation | Pegasus |
| Variability range | 18.1–19.5 *r*, 18.1–20.1 *g* |
| Orbital period | 6.5997 days (158.4 hours) |
| Epoch of mid-eclipse | JD=2459524.684 |
| Variability type | EA/WD |

Supposing the combined mass of two components in Grigoriev 1 to be about 1 solar mass, one can calculate the physical parameters of the binary system using 3$^{rd}$ Kepler's law. The orbital velocity can be easily computed by the formula $V=2\pi a/P$, where *a* – orbital radius, and *P* – orbital period determined from observations. The *V* value is equal to 113.5 km/s under the assumption of the mass of system being equal to 1 mass of the Sun.

Fig. 3 represents the part of the Grigoriev 1 phased plot near the eclipse in more detail. One can see the partial phases (ingress and egress) are about 30 times shorter than the main minimum. Based on this one can estimate the diameters of both components. They are given in Table 2.

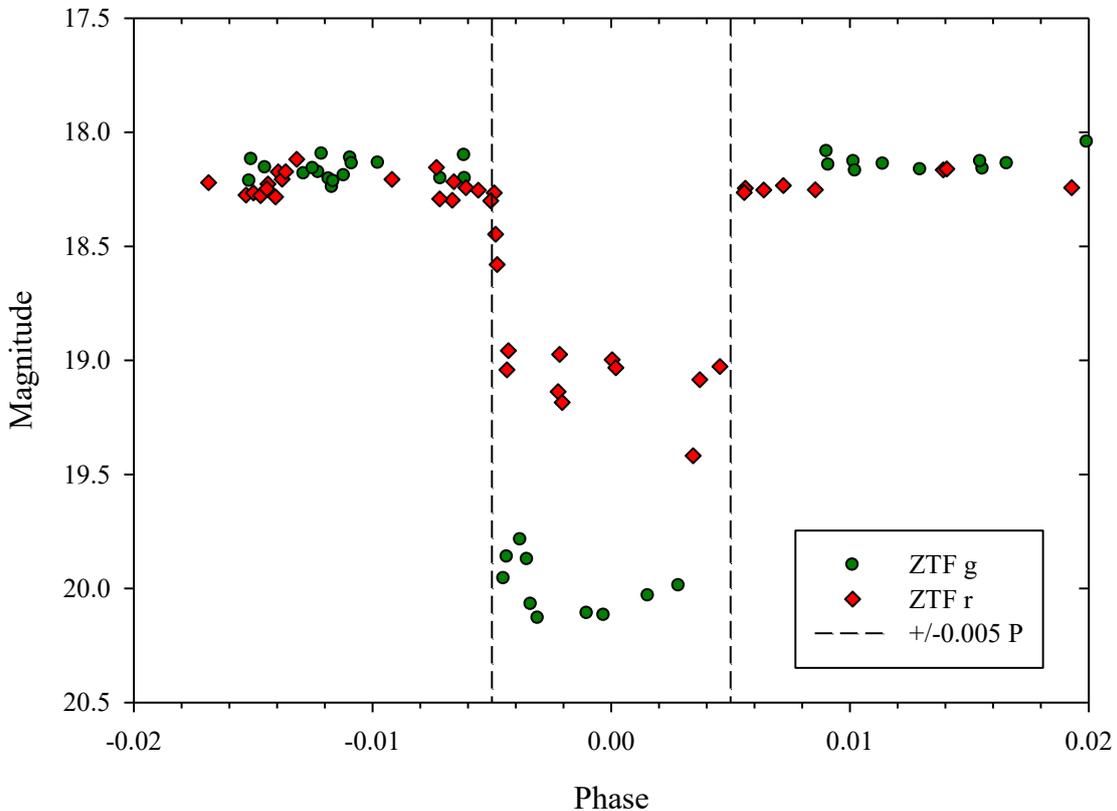

**Fig. 3.** Zoomed part of Grigoriev 1 phase plot in range of ±0.01 period

Table 2. Physical parameters of the binary system Grigoriev 1

| Parameter | Value |
|---|---|
| Orbital radius, AU | 0.06886 |
| Orbital radius, 1000 km | 10300 |
| Orbital velocity, km/s | 113.5 |
| Diameter of the cold component, 1000 km | 647 |
| Diameter of the hot component, 1000 km | ~20 |

As seen from Table 2, the radius of the cool companion is about 320 thousand km, or 0.032 orbital radius. That means the inclination of the orbital plane differs by less than 1.8° (0.032 rad) from 90 degrees. The probability of such orbital plane orientation in space is 3%. In other words, among 33 binary systems like Grigoriev 1 only one will show eclipses. This explains the fact that the eclipsing binaries like this have not been discovered before. Using the 3rd data release of Gaia space observatory (Gaia collaboration, 2022) one can derive the distance to Grigoriev 1 from its parallax 0.546±0.111 mas. It turns out to be 1830±370 parsec. The probability of finding other systems like this within 1 kpc is very small, and further on they are becoming fainter than 19–20 magnitude making them unavailable for discovery by the existing sky surveys.

Fig. 4 shows the distribution of EA/WD binary systems by the orbital periods. The size of the bin is 0.05 days. It is clearly seen that most of these systems are short-period ones ($P < 0.5$ days), with Grigoriev 1 standing out.

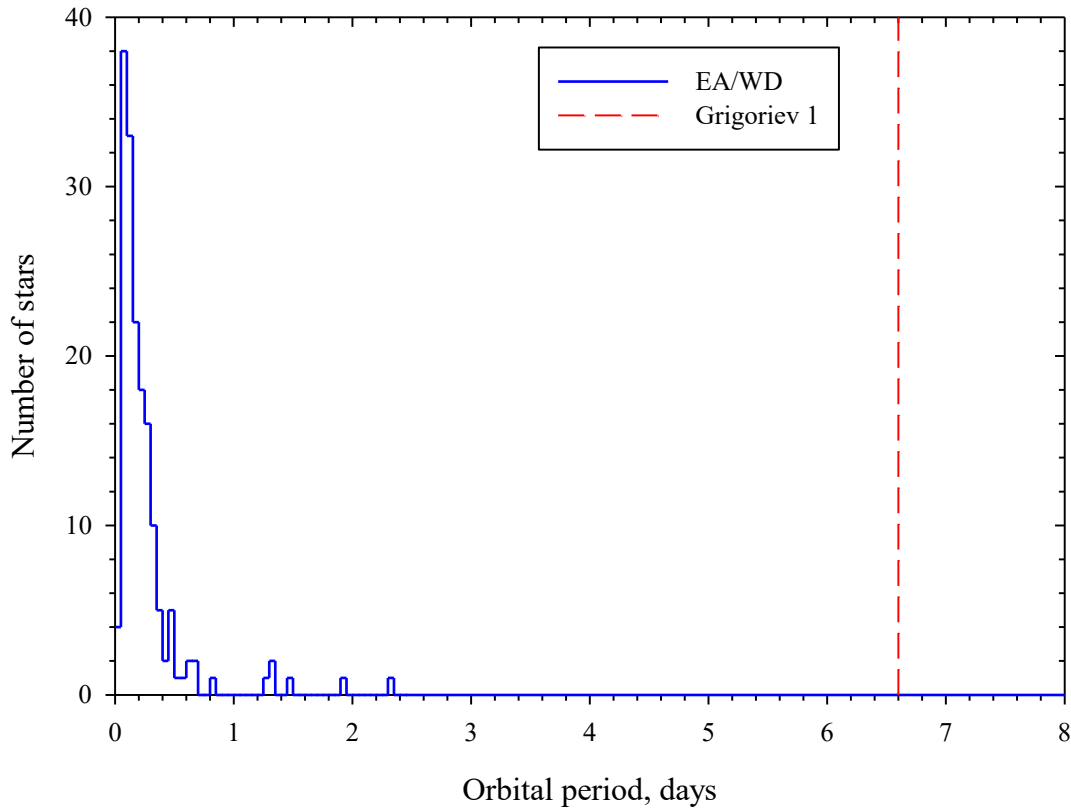

**Fig. 4.** Distribution of 188 known eclipsing binaries with white dwarfs by the orbital periods. The position of Grigoriev 1 is marked with the vertical red line.

The absolute magnitude of Grigoriev 1 at maximum light is +6.8. Fig. 5 shows the «color-absolute magnitude» diagram for the eclipsing variables with white dwarfs (EA/WD) and with hot subdwarfs of HW Vir type (EA/HW under VSX classification). Color indices (*B-R*) from Gaia DR3 are shown on the horizontal axis and absolute magnitudes in Gaia band ($M_G$) on the vertical axis. At this diagram Grigoriev 1 (marked with a blue star) occupies the intermediate position between the hot subdwarfs and white dwarfs which makes it even more remarkable. The hot component is most likely at the transition stage from one type of object to another. The color index in UV band from GALEX space observatory data (*FUV-NUV*=–0.666) is telling about the extremely high temperature of the hot component. It should be noted here that the reflection effect on the light curve is less than 0.1 magnitude because of a large distance between the components.

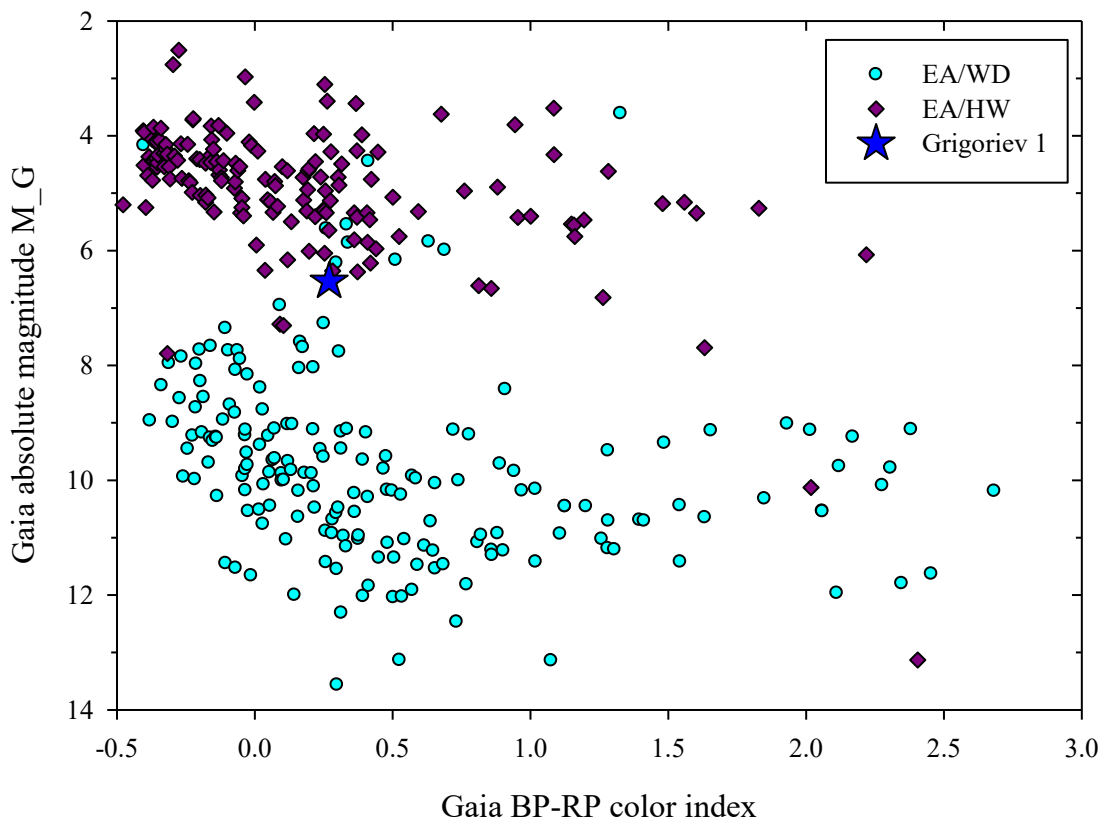

**Fig. 5.** Grigoriev 1 at the «color-luminosity» diagram for the eclipsing binaries with white dwarfs (blue circles) and hot subdwarfs (violet diamonds)

The star deserves further study with large telescopes. First, the radial velocity curve should be measured and the mass function determined. Rather slow orbital motion allows us to measure directly the sizes of components and the brightness distribution over the disk of hot subdwarf. During the eclipse lasting for an hour and a half one can obtain the spectrum of the cold companion and improve the mass, temperature and radius dependence of the red dwarfs in binary systems. Also the high-precision photometry near the secondary eclipse at the phases from 0.49 to 0.51 is of interest.

The work was performed in course of the project «Center of astronomical object discoveries» at the Moscow Palace of Pioneers (State education center "Vorobyovy Gory").